\documentstyle[12pt,epsf]{article}

\newcommand\ijmpa[3]{Int.\ J.\ Mod.\ Phys.\ A {\bf #1}, #3 (#2)}

\newcommand\plb[3]{Phys.\ Lett.\ B {\bf #1}, #3 (#2)} 
\newcommand\Prd[3]{Phys.\ Rev.\ D {\bf #1}, #3 (#2)}
\newcommand\Prl[3]{Phys.\ Rev.\ Lett.\ {\bf #1}, #3 (#2)}

\newcommand{\hepph}[1]{{\tt hep-ph/#1}}

\begin{document}

\begin{titlepage}

\begin{flushright}
CERN-TH/98-293\\
EFI-98-44\\
hep-ph/9809311
\end{flushright}

\vspace{1.0cm}
\begin{center}
\Large\bf 
\boldmath  
Determination of the Weak Phase $\gamma$ from\\
Rate Measurements in $B^\pm\to\pi K,\pi\pi$ Decays
\unboldmath
\end{center}

\vspace{0.5cm}
\begin{center}
Matthias Neubert\\[0.1cm] 
{\sl Theory Division, CERN, CH-1211 Geneva 23, Switzerland} \\[0.3cm] 
and\\[0.3cm] 
Jonathan L. Rosner\\[0.1cm]
{\sl Enrico Fermi Institute and Department of Physics\\ University of
Chicago, Chicago, IL 60637, USA}
\end{center}

\vspace{0.5cm}
\begin{abstract}
\vspace{0.2cm}
\noindent  
A method is described which, under the assumption of SU(3) symmetry,
allows one to determine the angle $\gamma=\mbox{Arg}(V_{ub}^*)$ of the
unitarity triangle from time-independent measurements of the
branching ratios for the rare two-body decays $B^+\to\pi^0 K^+$ and 
$B^-\to\pi^0 K^-$, as well as of the CP-averaged branching ratios for 
the decays $B^\pm\to\pi^\pm K^0$ and $B^\pm\to\pi^\pm\pi^0$, all of 
which are of order $10^{-5}$. The effects of electroweak penguin 
operators  are included in a model-independent way, and SU(3)-breaking  
corrections are accounted for in the factorization approximation.
\end{abstract}

\vspace{0.5cm}
\centerline{(To appear in Physical Review Letters)}

\vfil
\noindent
September 1998

\end{titlepage}

The study of CP violation in the weak decays of $B$ mesons will
provide important tests of the flavor sector of the Standard Model,
which predicts that all CP violation results from the presence of a
single complex phase in the quark mixing matrix. The precise
determination of the sides and angles of the unitarity triangle, which
is a graphical representation of the unitarity relation $V_{ub}^*
V_{ud}+V_{cb}^* V_{cd}+V_{tb}^* V_{td}=0$, plays a central role in this 
program \cite{review}. Whereas the angle $\beta=-\mbox{Arg}(V_{td})$ 
will be accessible at the first-generation $B$ factories through the 
measurement of CP violation in the decay $B\to J/\psi K_S$, the angle 
$\gamma=\mbox{Arg}(V_{ub}^*)$ is harder to determine. The sum 
$(\beta+\gamma)$ can be extracted in a theoretically clean way from 
measurements of CP violation in the decays $B\to\pi\pi$ (or in the 
related decays $B\to\pi\rho$ and $\rho\rho$), but because of 
experimental difficulties such as the detection of the mode 
$B\to\pi^0\pi^0$ this will be a long-term objective. A method to 
determine $\gamma$ proposed by Gronau and Wyler uses rate measurements 
for six $B\to D K$ decay modes \cite{GrWy}, some of which require the 
reconstruction of the neutral charm-meson CP eigenstate $D_+^0$. A 
variant of this approach using $B\to D K^*$ decays has been discussed 
by Dunietz \cite{Isi}. Unfortunately, these methods rely either on 
measurements of some processes with very small branching ratios, 
posing experimental \cite{Stone} and theoretical \cite{ADS} challenges, 
or on measurements requiring considerable precision (see, e.g.,
Refs.~\cite{NewDK,Jang} and references therein).

In view of these difficulties, approximate methods to determine the
angle $\gamma$ have received a lot of attention. The simplest of these
methods was proposed by Gronau, Rosner and London (GRL), who
suggested a triangle construction involving the amplitudes for the
decays $B^+\to\pi^+ K^0$, $\pi^0 K^+$, and $\pi^+\pi^0$, as well as 
for the corresponding CP-conjugated decays \cite{GRL}.  Besides a
plausible dynamical assumption this method relies on  SU(3) flavor
symmetry in relating $B\to\pi\pi$ with $B\to\pi K$ decays.  Later, it
was argued that the GRL method is spoiled by  electroweak penguin
contributions, which have an important impact  in $B\to\pi K$ decays
and upset the naive SU(3) triangle constructions
\cite{DeHe,GHLR2}. More sophisticated methods based on quadrangle
constructions involving other decay modes such as $B_s^0\to\pi^0\eta$
\cite{GHLR2} or $B^+\to\eta^{(\prime)} K^+$ \cite{DeHe2,GR96} were 
invented to circumvent this problem. There have also been
proposals for deriving bounds on $\gamma$ using CP-averaged rate
measurements in $B\to\pi K$ decays \cite{FM,Ge97,Ne97,Fa97,At97,us}, 
and for combining these measurements with those of rate asymmetries 
and other decays like $B\to K\bar K$ to obtain further information 
\cite{GRKpi,RFnew}.

In the present note we propose a variant of the original GRL method,
which based on the findings of our previous work \cite{us} includes
the potentially dangerous electroweak penguin  contributions in a
model-independent way using Fierz identities and SU(3) symmetry. We
thus obtain an approximate method for learning $\cos\gamma$ that is
conceptually as simple and uses the same experimental input and
theoretical assumptions as the GRL method, though the actual triangle
constructions are somewhat more complicated. The main advantage 
of our approach is that it is based on rare two-body decays that 
are relatively easy to access experimentally, and that have larger 
branching ratios than the decays needed for all other methods of 
measuring $\gamma$. Although the accuracy of this extraction may 
ultimately be limited by theoretical uncertainties, even an 
approximate value for $\cos\gamma$ will be very useful, if only to 
help eliminating discrete ambiguities inherent in other 
determinations \cite{Wolfen}.

The basis of our method is the amplitude relation
\begin{eqnarray}
   3 A_{3/2} &=& {\mathcal A}(B^+\to\pi^+ K^0)
    + \sqrt 2\,{\mathcal A}(B^+\to\pi^0 K^+) \nonumber\\
   &\approx& \sqrt 2\,\frac{V_{us}}{V_{ud}} \frac{f_K}{f_\pi}\,
    |{\mathcal A}(B^+\to\pi^+\pi^0)|\,e^{i\phi_{3/2}} 
    (\delta_{\rm EW} - e^{i\gamma}) \,, 
\label{SU3rel}
\end{eqnarray}
where $A_{3/2}$ is an isospin amplitude parametrizing the  $\Delta
I=1$ transition $B\to(\pi K)_{I=3/2}$, $e^{i\phi_{3/2}}$ is a
strong-interaction phase, and $e^{i\gamma}$ is the weak phase
associated with the quark decay $\bar b\to\bar u u\bar s$. The second
relation is strictly valid in the SU(3) flavor-symmetry limit;
however, the factor  $f_K/f_\pi=1.22\pm 0.01$ accounts for the leading
(i.e., factorizable)  corrections to that limit. The crucial new
ingredient in (\ref{SU3rel}) with respect to the corresponding
relation used in Ref.~\cite{GRL} is the presence of the parameter
$\delta_{\rm EW}$ accounting for the  contributions of electroweak
penguin operators. We have recently shown that in the SU(3) limit this
parameter is real (i.e., it does not carry a non-trivial
strong-interaction phase) and calculable  in terms of Wilson
coefficients and electroweak parameters \cite{us}. The result is
\begin{equation}
   \delta_{\rm EW} = (1-\kappa)\,\frac{1.71\alpha}{\lambda^2 R_b} =
   0.63\pm 0.11 \,,
\label{dEW}
\end{equation}
where $\alpha=1/129$ is the electromagnetic coupling at the weak
scale, $\lambda=0.22$ is the Wolfenstein parameter,
$R_b=\lambda^{-1}|V_{ub}/V_{cb}|\approx 0.41\pm 0.07$ \cite{Rosnet},
and $\kappa\approx 0.05$ accounts for factorizable SU(3)-breaking
corrections. The derivation of this result uses the fact that the
relevant electroweak penguin operators are Fierz-equivalent to the
usual current--current operators $Q_1$ and $Q_2$ of the effective weak
Hamiltonian for $B\to\pi K$ decays \cite{Ne97}, and that in  the SU(3)
limit the isospin amplitude $A_{3/2}$ receives a  contribution only
from the combination $(Q_1+Q_2)$, but not from the  difference
$(Q_1-Q_2)$.

As in the original GRL method, we must rely on the dynamical
assumptions that $|{\mathcal A}(B^+\to\pi^+\pi^0)| = |{\mathcal
A}(B^-\to\pi^-\pi^0)|$ and ${\mathcal A}(B^+\to\pi^+ K^0) =
{\mathcal A}(B^-\to\pi^-\bar K^0)$. Whereas the first  relation
follows from the fact that only the current--current  operators
contribute to $B^\pm\to\pi^\pm\pi^0$ decays (electroweak  penguin
contributions can be neglected in this case \cite{Flei96}),  the
second one assumes that there are only negligible contributions
proportional to the weak phase $e^{i\gamma}$ to the amplitude for the
decay $B^+\to\pi^+ K^0$, which thus can be taken to have the simple
form ${\mathcal A}(B^+\to\pi^+ K^0) = e^{i\pi} e^{i\phi_P}  |{\mathcal
A}(B^+\to\pi^+ K^0)|$, where $e^{i\pi}$ is the weak  phase of the
leading top- and charm-penguin amplitudes, and  $e^{i\phi_P}$ is a
strong-interaction phase. Possible contributions to this amplitude
proportional to the weak phase $e^{i\gamma}$ are indeed expected to 
be very small, because they could only come from  up-quark penguins or
annihilation topologies \cite{ampl}. However, this intuitive argument
could be invalidated if soft final-state  rescattering effects were
very important \cite{Ge97,Ne97,Fa97,At97,GRKpi,RFnew}. We stress,
therefore, that the assumption  ${\mathcal A}(B^+\to\pi^+ K^0) =
{\mathcal A}(B^-\to\pi^-\bar K^0)$ is a working hypothesis of our
method, which must be tested independently. A necessary condition for
the validity of this assumption is the absence of a sizable direct CP
asymmetry in the decays  $B^\pm\to\pi^\pm K^0$. If we write ${\mathcal
A}(B^+\to\pi^+ K^0) \propto e^{i\phi_P}(e^{i\pi}+e^{i\gamma} e^{i\eta}
\varepsilon_a)$, where $\varepsilon_a\ll 1$ measures the strength of 
possible rescattering contributions and $e^{i\eta}$ is a 
strong-interaction phase, then $a_{\rm CP}\approx 2\varepsilon_a 
\sin\gamma\sin\eta$. Since the global analysis of the unitarity 
triangle prefers values of $\gamma$ such that $\sin\gamma=O(1)$, and 
since $\sin\eta$ is unlikely to be small because without sizable 
strong phases there would not be a rescattering contribution in the 
first place, a small experimental value for the asymmetry would be a 
strong indication that our working hypothesis is justified.

Let us define the amplitude ratios
\begin{eqnarray}
   \varepsilon_{3/2} &=& \frac{V_{us}}{V_{ud}} \frac{f_K}{f_\pi}
    \frac{\sqrt 2\,|{\mathcal A}(B^+\to\pi^+\pi^0)|}
         {|{\mathcal A}(B^+\to\pi^+ K^0)|} \,, \nonumber\\
   r_\pm &=& \frac{\sqrt 2\,|{\mathcal A}(B^\pm\to\pi^0 K^\pm)|}
                  {|{\mathcal A}(B^+\to\pi^+ K^0)|} \,,
\end{eqnarray}
which under the assumptions stated above can be determined 
experimentally through time-independent rate measurements via
\begin{eqnarray}
   \varepsilon_{3/2}\! &=& \sqrt 2\,\frac{V_{us}}{V_{ud}}
    \frac{f_K}{f_\pi} \left[ \frac{\mbox{Br}(B^+\to\pi^+\pi^0) +
    \mbox{Br}(B^-\to\pi^-\pi^0)} {\mbox{Br}(B^+\to\pi^+ K^0) +
    \mbox{Br}(B^-\to\pi^-\bar K^0)} \right]^{1/2} , \nonumber\\
   r_\pm &=& 2 \left[ \frac{\mbox{Br}(B^\pm\to\pi^0 K^\pm)}
    {\mbox{Br}(B^+\to\pi^+ K^0) + \mbox{Br}(B^-\to\pi^-\bar K^0)}
    \right]^{1/2} .
\end{eqnarray}
A future measurement of $r_+\ne r_-$ would signal direct CP violation 
in the decays $B^\pm\to\pi^0 K^\pm$. At present, preliminary data 
reported by the CLEO Collaboration \cite{CLEOnew} imply 
$[\frac 12(r_+^2 + r_-^2)]^{1/2}=1.46\pm 0.37$ and, combined with some 
theoretical guidance, $\varepsilon_{3/2}=0.24\pm 0.06$ \cite{us}. 
Moreover, we define
\begin{equation}
   \delta_{\rm EW} - e^{i\gamma}\equiv \varrho(z)\,e^{-i\psi}
\end{equation}
with $z=\cos\gamma$, so that
\begin{equation}
   \varrho(z) = \sqrt{1 - 2z\delta_{\rm EW} + \delta_{\rm EW}^2} \,,
   \qquad \sin\psi = \frac{\sin\gamma}{\varrho(z)} \,.
\end{equation}
In terms of these quantities, the triangle relation (\ref{SU3rel})
and its CP-conjugate take the form
\begin{equation}
   1 + \varepsilon_{3/2}\varrho(z)\,e^{i(\Delta\phi\mp\psi)} =
   r_\pm\,e^{i\xi_\pm} \,,
\label{triangle}
\end{equation}
where $\Delta\phi=\phi_{3/2}-\phi_P$ is an unknown strong-interaction
phase difference, while the phases $\xi_\pm$ contain both strong and
weak contributions. It follows that
\begin{eqnarray}
   \cos(\psi\mp\Delta\phi) &=&
    \frac{r_\pm^2-1-\varepsilon_{3/2}^2\varrho^2(z)}
     {2\varepsilon_{3/2}\varrho(z)} \equiv x_\pm(z) \,, \nonumber\\
    \cos(2\psi) &=& 1 - \frac{2(1-z^2)}{\varrho^2(z)} \,.
\end{eqnarray}
Combining these results, we find that the allowed solutions for
$z=\cos\gamma$ can be obtained from the real zeros of the equation
\begin{equation}
   \frac{\big(r_+^2 - r_-^2\big)^2}{16\varepsilon_{3/2}^2} +
   \frac{\big(1-z^2\big)^2}{\varrho^2(z)} = (1-z^2) \Big[ 1 -
   x_+(z)\,x_-(z) \Big] \,,
\label{poly}
\end{equation}
which, taking into account the $z$ dependence of $\varrho(z)$ and
$x_\pm(z)$, correspond to the zeros of a fourth-order polynomial in
$z$.

A simplified analysis can be performed if the phase difference
$\Delta\phi$ turns out to be small or close to $180^\circ$ -- 
a possibility that can be tested for experimentally. To this end, 
one exploits the following exact relations:
\begin{eqnarray}
   \cos\gamma &=& \delta_{\rm EW}
    - \frac{\frac 12(r_+^2 + r_-^2)-1-\varepsilon_{3/2}^2
            (1-\delta_{\rm EW}^2)}
           {2\varepsilon_{3/2}(\cos\Delta\phi
            +\varepsilon_{3/2}\delta_{\rm EW})} \,, \nonumber\\
   r_+^2 - r_-^2 &=& 4\varepsilon_{3/2}\sin\gamma\sin\Delta\phi \,.
\label{simple}
\end{eqnarray}
The global analysis of the unitarity triangle prefers values of
$\gamma$ in the range $47^\circ<\gamma<105^\circ$ \cite{Jonnew}, which
would imply $\sin\gamma>0.73$. Then the second relation can be used to
obtain a reasonable estimate and upper limit for $\sin\Delta\phi$. If it 
turns out that $\sin\Delta\phi$ is small, corresponding to a situation 
where $|\Delta\phi|\approx 0^\circ$ or $180^\circ$, one can set
$\cos\Delta\phi=\pm 1$ in the first relation to obtain
\begin{equation}
   \cos\gamma \approx 
   \frac{(1\pm\varepsilon_{3/2}\delta_{\rm EW})^2
         - \frac 12(r_+^2 + r_-^2) + \varepsilon_{3/2}^2}
    {2\varepsilon_{3/2}(\pm 1+\varepsilon_{3/2}\delta_{\rm EW})} \,,
\label{cute}
\end{equation}
which determines $\cos\gamma$ up to a possible two-fold ambiguity.
From (\ref{simple}), it follows that a criterion for the validity of 
this approximation is that the deviation of $\cos\Delta\phi$ from 
$\pm 1$ be less than the uncertainty in the product $\varepsilon_{3/2}
\delta_{\rm EW}$, i.e.\ $\mbox{min}(|\Delta\phi|,|\Delta\phi-\pi|)
<\sqrt{2\Delta(\varepsilon_{3/2}\delta_{\rm EW})}$. With present 
uncertainties on the parameters $\varepsilon_{3/2}$ and 
$\delta_{\rm EW}$, which are unlikely to be improved much in the near 
future, this implies $\mbox{min}(|\Delta\phi|,|\Delta\phi-180^\circ|)
<17^\circ$. With the current experimental values for the various 
parameters, and in the absence of independent experimental results for 
$r_+$ and $r_-$, the relations (\ref{simple}) do not yet provide for a 
useful estimate of $\cos\gamma$; however, they may become valuable 
with more precise measurements. It is remarkable that even in the case 
$r_+=r_-$, i.e., in the absence of direct CP violation in 
$B^\pm\to\pi^0 K^\pm$ decays, $\cos\gamma$ can be determined using 
relation (\ref{cute}), which becomes exact in that limit.

\begin{figure}
\epsfxsize=8cm  
\centerline{\epsffile{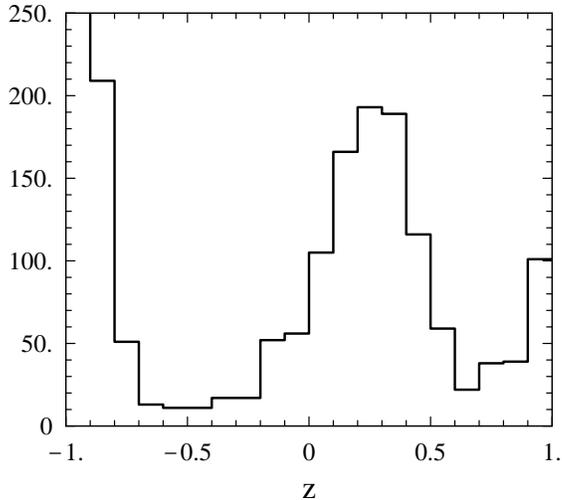}}
\vspace{0.2cm}
\centerline{\parbox{14cm}{\caption{\sl 
Real solutions for $z=\cos\gamma$ obtained from (\protect\ref{poly}) in 
a simulation of 1000 experiments with Gaussian errors as specified in 
the text. The correct value is $z=\cos 76^\circ\approx 0.242$.}}}
\label{fig:scatter}
\end{figure}

In practise, the determination of $\gamma$ using (\ref{poly}) or
(\ref{simple}) is limited by experimental as well as  theoretical
uncertainties in the extraction of the parameters $r_\pm$,
$\varepsilon_{3/2}$, and $\delta_{\rm EW}$. Let us illustrate the
situation with a realistic example. Assume that the true values of the
parameters are $\gamma=76^\circ$ (the center of the region preferred
by the global analysis), $\varepsilon_{3/2}=0.24$ and $\delta_{\rm
EW}=0.63$ (the current central values), and that the strong phase
difference takes the value $\Delta\phi=20^\circ$. It then follows that
$r_+\approx 1.18$ and $r_-\approx 1.04$.  Let us assume that we can
measure the values of these parameters with some errors given by
$\Delta\varepsilon_{3/2}=0.04$, $\Delta\delta_{\rm EW}=0.09$, and
$\Delta r_\pm=0.05$.  We do not anticipate that $\varepsilon_{3/2}$
and $\delta_{\rm EW}$ will soon be known with an accuracy much better
than today, because these quantities are affected by theoretical
uncertainties such as the estimate of SU(3)-breaking effects.  We thus
assign a 15\% error to   them \cite{eps}.  The assumed error on the
amplitude ratios $r_\pm$ corresponds to a measurement of the
corresponding ratios of branching ratios  with a precision of about
10\%.  In this example, the approximate value for $\cos\gamma$
obtained by setting $\cos\Delta\phi=1$ in (\ref{simple}) is
$\cos\gamma\approx 0.26\pm 0.14(r_\pm)\pm 0.09(\delta_{\rm EW})\pm
0.09(\varepsilon_{3/2})$, which is close to the correct value
$\cos\gamma\approx 0.242$. We have quoted the various sources of
errors separately. It is apparent that the precision in the
measurements of the ratios  $r_\pm$ is the limiting factor of our
method. The approximate solution obtained with 
$\cos\Delta\phi=-1$ is $\cos\gamma\approx 1.13\pm 0.19(r_\pm)\pm 
0.12(\delta_{\rm EW})\pm 0.10(\varepsilon_{3/2})$, which is excluded 
by the global analysis of the unitarity triangle. In 
Figure~\ref{fig:scatter}, we show the distribution of the exact real 
solutions of equation (\ref{poly}) for 1000 random choices of the 
Gaussian errors in the various input quantities. The solutions where 
$\cos\gamma\approx\pm 1$ can again be excluded based on the global 
analysis of the unitarity triangle. From the central peak, we obtain 
$\cos\gamma=0.24\pm 0.18$, implying at one standard deviation 
$|\gamma|=(76\pm 11)^\circ$.

To conclude, we have shown that the weak phase
$\gamma=\mbox{Arg}(V^*_{ub})$ can be determined using
time-independent measurements of the branching ratios for the decays
$B^+\to\pi^0 K^+$ and $B^-\to\pi^0 K^-$, as well as of the CP-averaged
branching ratios for the decays $B^\pm\to\pi^\pm K^0$ and
$B^\pm\to\pi^\pm\pi^0$.  The new development that makes this method
practical is the observation that the strong phases of the $I=\frac
32$ electroweak penguin and tree amplitudes are related to one another
by Fierz identities and SU(3) flavor symmetry. SU(3)-breaking 
corrections can be accounted for in the factorization approximation. 
On the other hand, like many earlier
proposals our method relies on the dynamical assumption that 
final-state rescatterings do not induce a sizable contribution 
proportional to the weak phase $e^{i\gamma}$ in the amplitude for the 
process $B^+\to\pi^+ K^0$. The validity of this assumption can be 
tested for experimentally by searching for direct CP violation in 
this decay.

\vspace{0.15cm}  
{\it Acknowledgments:\/} 
Part of this work was done during the Workshop on {\sl Perturbative 
and Non-Perturbative Aspects of the Standard Model\/} at St.\ John's 
College, Santa Fe, July--August 1998. We would like to thank the 
organizer Rajan Gupta, as well as the participants of the workshop, 
for providing a stimulating atmosphere and for many useful discussions.  
We also wish to thank Michael Gronau, Sheldon Stone and Lincoln 
Wolfenstein for helpful comments. One of us (J.L.R.) was supported in 
part by the United States  Department of Energy through contract No.\ 
DE FG02 90ER40560.


\begin{thebibliography}{99}
\bibitem{eps}   
One may be able to reduce the error on $\varepsilon_{3/2}$ somewhat by   
studies of factorization in $B\to\pi^+\pi^-$ and $B\to\pi\,\ell\,  
\bar\nu_\ell$ decays; see the first paper in Ref.~\protect\cite{GRKpi},  
where an ultimate error of 10\% was projected for a related quantity  
without the color-suppressed contribution.
\end{thebibliography}
\end{document}